\documentclass{elsart}

\usepackage{graphicx}
\usepackage{amssymb}

\begin{document}

\begin{frontmatter}

\title{Universality of strength for Yukawa couplings
with extra down-type quark singlets}

\author[Kaisei]{Katsuichi Higuchi},
\author[ICRR]{Masato Senami},
\author[Kyoto]{Katsuji Yamamoto\corauthref{cor}}
\corauth[cor]{Corresponding author.}
\ead{yamamoto@nucleng.kyoto-u.ac.jp}

\address[Kaisei]{Department of Literature, Kobe Kaisei College,
Kobe 657-0805, Japan}
\address[ICRR]{Institute for Cosmic Ray Research, University of Tokyo,
Kashiwa, Chiba 277-8582, Japan}
\address[Kyoto]{Department of Nuclear Engineering, Kyoto University,
Kyoto 606-8501, Japan}

\begin{abstract}
We investigate the quark masses and mixings
by including vector-like down-type quark singlets
in universality of strength for Yukawa couplings (USY).
In contrast with the standard model with USY,
the sufficient $ CP $ violation is obtained
for the Cabibbo-Kobayashi-Maskawa matrix
through the mixing between the ordinary quarks and quark singlets.
The top-bottom mass hierarchy $ m_t \gg m_b $
also appears naturally in the USY scheme with the down-type quark singlets.
\end{abstract}

\end{frontmatter}

The explanation of the masses and mixings of quarks and leptons
is one of the fundamental issues in particle physics.
Many notable ideas to address this problem have been investigated,
including the universality of strength for Yukawa couplings (USY)
\cite{USY,PPMM}.
In the standard model with USY, the nearly democratic quark mass matrices
\cite{DMM} are provided, and the quark masses
and the magnitude of the Cabibbo-Kobayashi-Maskawa (CKM) matrix
are really reproduced with the suitable USY phases.
However, the USY scheme seems to confront some difficulties
within the context of the standard model.
Some reasonable explanation should be presented
for the top-bottom mass hierarchy $ m_t \gg m_b $;
it is simply attributed to the hierarchy of the Yukawa couplings
between the up and down sectors with one Higgs doublet,
or a large ratio of the vacuum expectation values (VEV's)
of two Higgs doublets.
More seriously, it is quite difficult
to obtain the sufficient $ CP $ violation for the CKM matrix
in the standard model with USY
\cite{USY,SUSY-USY},
which is essentially due to the fact that the USY phases
are small to provide the quark masses except for the third generation.

In this letter, we present a new look at the USY scheme
by including exotic ingredients.
Specifically, we investigate an extension of the standard model
with extra down-type quark singlets
\cite{quark-singlets,CP-EQ1,CP-EQ2,HY}.
The standard model contains three generations of the ordinary quarks,
left-handed doublets
$ q_{i{\rm L}} = ( u_{i{\rm L}} , d_{i{\rm L}} )^{\rm T} $
and right-handed singlets $ u_{i{\rm R}} $, $ d_{i{\rm R}} $
($ i = 1,2,3 $), and a Higgs doublet $ H $.
In addition, $ N_D $ vector-like down-type quark singlets
$ D_{a{\rm L}} $ and $ D_{a{\rm R}} $
($ a = 4, \ldots , 3 + N_D $) and a Higgs singlet $ S $ are included
\cite{CP-EQ2,HY}, which may be accommodated in E6-type models.
We will show that the actual quark masses and CKM matrix
are indeed obtained in the USY scheme with extra down-type quark singlets.
In particular, through the $ d $-$ D $ mixing
the sufficient $ CP $ violation for the CKM matrix is provided
from some large USY phases of the Yukawa couplings
with the Higgs singlet $ S $.
(This mixing mechanism to transmit the $ CP $ violation
is considered in Refs. \cite{CP-EQ1,CP-EQ2}.)
The top-bottom hierarchy $ m_t \gg m_b $ also appears naturally
in the USY scheme (or more generally flavor democracy)
due to the existence of extra down-type quark singlets
but no such up-type quark singlets as in the {\bf 27} of E6.

The Yukawa couplings of quarks and Higgs fields with USY
are given by
\begin{eqnarray}
{\mathcal L}_{\rm Y}
&=& - 
{\bar q}_{i{\rm L}} \Lambda^u_{ij} u_{j{\rm R}} H
- {\bar q}_{i{\rm L}} \Lambda^d_{iJ} {\mathcal D}_{J{\rm R}}
( i \sigma_2 H^* )
\nonumber \\
& & - {\bar D}_{a{\rm L}} \Lambda^D_{aJ} {\mathcal D}_{J{\rm R}} S
+ {\rm H.c.} ,
\nonumber \\
\Lambda^u_{ij} &=& \frac{\lambda_u}{3} {\rm e}^{i \phi^u_{ij}} ,
\Lambda^d_{iJ} = \frac{\lambda_d}{3} {\rm e}^{i \phi^d_{iJ}} ,
\Lambda^D_{aJ} = \frac{\lambda_D}{3} {\rm e}^{i \phi^D_{aJ}} ,
\label{Yukawa-USY}
\end{eqnarray}
where $ J = j , b $
with $ {\mathcal D}_j \equiv d_j $ and $ {\mathcal D}_b \equiv D_b $.
The respective types of Yukawa couplings are specified
with the strengths $ \lambda_u $, $ \lambda_d $, $ \lambda_D $
and USY phases $ \phi^u_{ij} $, $ \phi^d_{iJ} $, $ \phi^D_{aJ} $.
The couplings $ {\bar D}_{a{\rm L}} {\mathcal D}_{J{\rm R}} S^* $
are excluded here for definiteness if $ S $ is a complex field.
This is really the case for the supersymmetric model
with a pair of Higgs doublets.
The quark mass matrices are given from Eq. (\ref{Yukawa-USY}) as
\begin{eqnarray}
M_u = ( \lambda v / 3 ) \left( {\rm e}^{i \phi^u_{ij}} \right) ,
{\mathcal M}_{\mathcal D} = ( \lambda v / 3 ) \left( \begin{array}{c}
{\rm e}^{i \phi^d_{iJ}} \\ \kappa {\rm e}^{i \phi^D_{aJ}}
\end{array} \right) ,
\end{eqnarray}
where $ \langle H^0 \rangle = v $, $ \langle S \rangle = v_S $
(the possible phase is absorbed by $ \phi^D_{aJ} $),
and $ \kappa = v_S / v $.
We investigate the case of $ \lambda_u = \lambda_d = \lambda_D = \lambda $
for definiteness, while the result is readily extended
for different $ \lambda_u $, $ \lambda_d $, $ \lambda_D $.

We first consider the up-type quark mass matrix
\begin{eqnarray}
M_u = M_u (0) + \Delta M_u
= \left( \begin{array}{ccc}
1 & 1 & 1 \\ 1 & 1 & 1 \\ 1 & 1 & 1 \end{array} \right) + \Delta M_u ,
\label{Mu}
\end{eqnarray}
where the perturbation part is given as $ \Delta M_u \simeq i \Phi_u $
with the small USY phase matrix $ ( \Phi_u )_{ij} = \phi^u_{ij} $.
Henceforth the quark mass terms are presented
to be dimensionless measured
in unit of $ \lambda v / 3 $ ($ \simeq m_t / 3 $).
The up-type quark mass matrix is relevantly expressed
in the hierarchical basis by making a suitable transformation
\cite{USY,PPMM}:
\begin{eqnarray}
{\widetilde M}_u
= U_q^\dagger M_u U_q I_u
\simeq {\rm diag} ( 0 , 0 , 3 ) + i {\widetilde \Phi}_u I_u ,
\label{Mu-hie}
\end{eqnarray}
where $ {\widetilde \Phi}_u = U_q^\dagger \Phi_u U_q $,
$ U_q = U(1)_{[12]} U({\sqrt 2})_{[23]} $,
$ I_u = {\rm diag} ( - i , - i , 1 ) $,
and ``$ {\rm diag} $" represents a diagonal matrix.
The unitary transformation $ U( \alpha )_{[IJ]} $
between the $ I $-th and $ J $-th quarks
is specified with a $ 2 \times 2 $ matrix
\begin{eqnarray}
U( \alpha ) = \frac{1}{\sqrt{ 1 + | \alpha |^2 }}
\left( \begin{array}{cc} 1 & \alpha \\ - \alpha^* & 1 \end{array} \right) ,
\nonumber
\end{eqnarray}
supplemented with the right dimension, $ 3 \times 3 $ for the up sector
and $ ( 3 + N_D ) \times ( 3 + N_D ) $ for the down sector.
We note here that by suitably choosing the phases
of $ q_{i{\rm L}} $'s and $ u_{j{\rm R}} $'s,
the USY phases are taken in general as
$ \phi^u_{i3} = - \phi^u_{i1} - \phi^u_{i2} $
and $ \phi^u_{3j} = - \phi^u_{1j} - \phi^u_{2j} $,
giving $ ( {\widetilde \Phi}_u )_{i3} = ( {\widetilde \Phi}_u )_{3j} = 0 $.
In this USY phase convention,
the pre-factor $ i $ for $ {\tilde \Phi}_u $
is practically removed with $ I_u $,
and the up-type quark mass matrix in Eq. (\ref{Mu-hie}) is given as
\begin{eqnarray}
{\widetilde M}_u
= V_{u{\rm L}} {\rm diag} ( m_u , m_c , m_t ) V_{u{\rm R}}^\dagger
\simeq \left( \begin{array}{ccc}
{\tilde \phi}_{u1} & {\tilde \phi}_{u2} & 0 \\
{\tilde \phi}_{c1} & {\tilde \phi}_{c2} & 0 \\
0 & 0 & 3 \end{array} \right)
\label{Mu-Phiu}
\end{eqnarray}
with $ {\tilde \phi}_{u_i j} = ( {\widetilde \Phi}_u )_{ij} $
($ u_1 = u , u_2 = c , u_3 = t $).

The quark mass hierarchy $ m_u \ll m_c \ll m_t $
for the nearly democratic $ M_u $ in Eq. (\ref{Mu})
is understood in terms of the sequential breakings
of the permutation symmetry $ S^3_{q{\rm L}} $
among the left-handed quark doublets \cite{S3}:
\begin{eqnarray}
S^3_{q{\rm L}} \rightarrow S^2_{q{\rm L}} \rightarrow {\mbox{\rm non}} .
\label{S3qL-S2qL}
\end{eqnarray}
The democratic and $ S^3_{q{\rm L}} $ invariant $ M_u (0) $
provides the top mass.
Then, for the USY phases in Eq. (\ref{Mu-Phiu})
the $ S^2_{q{\rm L}} $ invariant terms $ {\tilde \phi}_{cj} $
and the small $ S^2_{q{\rm L}} $ breaking ones $ {\tilde \phi}_{uj} $
provide the charm and up masses, respectively, as
\begin{eqnarray}
| {\tilde \phi}_{cj} | \sim m_c \gg | {\tilde \phi}_{uj} | \sim m_u
\label{phiu-tilde}
\end{eqnarray}
with $ ( V_{u{\rm L}} )_{12} \sim m_u / m_c \ll 1 $
($ V_{u{\rm L}} \simeq {\bf 1} $).

We next investigate the down sector including two singlet $ D $'s,
while the essential features are valid for $ N_D \geq 2 $.
The USY scheme with only one $ D $ is, however, unsatisfactory,
still providing the too small $ CP $ violation for the CKM matrix.
This is because the USY phases $ \phi^D_{4J} $ in $ \Lambda^D $
with the Higgs singlet $ S $ are all eliminated away
by rephasing $ {\mathcal D}_{J{\rm R}} $'s.
Then, the remaining USY phases should be small
to provide the ordinary quark masses
just as in the standard model with USY.

The down-type quark mass matrix is given as
\begin{eqnarray}
{\mathcal M}_{\mathcal D}
= {\mathcal M}_{\mathcal D} (0) + \Delta {\mathcal M}_{\mathcal D} .
\end{eqnarray}
The main part has a quasi-democratic form
\begin{eqnarray}
{\mathcal M}_{\mathcal D} (0)
= \left( \begin{array}{ccccc}
1 & 1 & 1 & 1 & 1 \\
1 & 1 & 1 & 1 & 1 \\
1 & 1 & 1 & 1 & 1 \\
\kappa & \kappa & \kappa & \kappa & \kappa \\
\kappa & \kappa & \kappa & \kappa & \kappa
\end{array} \right)
\end{eqnarray}
with $ \kappa = v_S / v $.
The remaining part $ \Delta {\mathcal M}_{\mathcal D} $
is provided with the USY phase matrix $ \Phi_{\mathcal D} $,
$ ( \Phi_{\mathcal D} )_{iJ} = \phi^d_{iJ} $
and $ ( \Phi_{\mathcal D} )_{aJ} = \phi^D_{aJ} $.
In accordance with Eq. (\ref{Mu-hie}) for the up sector,
the mass matrix of down sector is transformed as
\begin{eqnarray}
{\widetilde{\mathcal M}}_{\mathcal D}
&=& U(1)_{[45]}^\dagger U_q^\dagger {\mathcal M}_{\mathcal D}
U_q U({\sqrt 3})_{[34]} U(2)_{[45]} I_{\mathcal D}
\nonumber \\
&=& \left( \begin{array}{cc}
{\widetilde M}_d & {\widetilde \Delta}^\prime_{dD} \\
{\widetilde \Delta}_{dD} & {\widetilde M}_D \end{array} \right) ,
\label{MD-total}
\end{eqnarray}
where $ I_{\mathcal D}
= {\rm diag} ( - i , - i , - i , - {\rm e}^{- i \theta} , 1 ) $
with $ \theta $ to be fixed below in Eq. (\ref{Delta}).
The USY phase matrix $ \Phi_{\mathcal D} $ is transformed
in the same way to $ {\widetilde \Phi}_{\mathcal D} I_{\mathcal D} $.
This transformation respects
the $ {\rm SU}(2)_W \times {\rm U}(1)_Y $ gauge symmetry
without $ d_{\rm L} $-$ D_{\rm L} $ mixing.
The main part is given in this basis as
\begin{eqnarray}
{\widetilde{\mathcal M}}_{\mathcal D} (0)
= \left( \begin{array}{ccccc}
0 & 0 & 0 & 0 & 0 \\
0 & 0 & 0 & 0 & 0 \\
0 & 0 & 0 & 0 & {\sqrt{15}} \\
0 & 0 & 0 & 0 & 0 \\
0 & 0 & 0 & 0 & {\sqrt{10}} \kappa
\end{array} \right) ,
\end{eqnarray}
providing four ($ 3 + N_D - 1 $) zero eigenvalues.
Hence, in contrast with the flavor democracy in the standard model,
the bottom quark no longer acquires so a heavy mass as the top quark.
This reasonably explains the top-bottom hierarchy
$ m_t \gg m_b $ in the USY scheme.
It is also noticed that one ($ N_D - 1 $) $ D $
should obtain a mass from the USY phases as well as the ordinary $ d $'s.

The USY phases in $ \Lambda^d $ with the Higgs doublet $ H $
are supposed to be small
to provide the ordinary quark masses and mixings.
On the other hand, those in $ \Lambda^D $ with the Higgs singlet $ S $
may be large to provide the significant $ CP $ violation for the CKM matrix
through the $ d $-$ D $ mixing.
It is convenient here to make $ \phi^D_{5J} = 0 $
by rephasing $ {\mathcal D}_{J{\rm R}} $'s.
We may also take for simplicity
$ \phi^D_{4j} \approx \phi^D_{44} \approx 0 $
under the approximate $ S^4_{{\mathcal D}{\rm R}} $
among $ {\mathcal D}_{1{\rm R}} $ -- $ {\mathcal D}_{4{\rm R}} $
together with the rephasing of $ {\mathcal D}_{4{\rm L}} $
(though not essential for the desired $ CP $ violation).
That is, in this convention
\begin{eqnarray}
| \phi^d_{iJ} | , | \phi^D_{4j} | , | \phi^D_{44} | \ll 1 ,
| \phi^D_{45} | \sim 1 , \phi^D_{5J} = 0 .
\end{eqnarray}

The submatrix $ {\widetilde M}_D $
in $ {\widetilde{\mathcal M}}_{\mathcal D} $ is given as
\begin{eqnarray}
{\widetilde M}_D \simeq \frac{\kappa}{\sqrt{10}}
\left( \begin{array}{cc} 2 | \Delta | & \Delta \\
2 | \Delta | & 10 + \Delta \end{array} \right) ,
\end{eqnarray}
where
\begin{eqnarray}
\Delta \equiv | \Delta | {\rm e}^{i \theta}
\equiv \exp [ i \phi^D_{45} ] - 1
\label{Delta}
\end{eqnarray}
with
$ \theta = [( | \phi^D_{45} | + \pi ) / 2 ] {\rm Sign} [ \phi^D_{45} ] $.
Then, the masses of the heavy quarks, almost the singlets, are given as
\begin{eqnarray}
m_{D_1} \simeq ( 2 / {\sqrt{10}} ) | \Delta | \kappa ,
m_{D_2} \simeq {\sqrt{10}} \kappa .
\end{eqnarray}

The submatrix $ {\widetilde M}_d $ for the ordinary quarks is given as
\begin{eqnarray}
{\widetilde M}_d
&=& V^{(0)}_{d{\rm L}} {\rm diag} ( m_d^{(0)} , m_s^{(0)} , m_b^{(0)} )
V^{(0) \dagger}_{d{\rm R}}
\nonumber \\
& \simeq & {\widetilde \Phi}_{\mathcal D}^{(3)}
= \left( \begin{array}{ccc}
{\tilde \phi}_{d1} & {\tilde \phi}_{d2} & {\tilde \phi}_{d3} \\
{\tilde \phi}_{s1} & {\tilde \phi}_{s2} & {\tilde \phi}_{s3} \\
{\tilde \phi}_{b1} & {\tilde \phi}_{b2} & {\tilde \phi}_{b3}
\end{array} \right)
\label{Md-Phid}
\end{eqnarray}
with $ m_{d_i}^{(0)} \sim m_{d_i} $ ($ d_1 = d , d_2 = s , d_3 = b $),
where $ {\widetilde \Phi}_{\mathcal D}^{(3)} $
is the $ 3 \times 3 $ submatrix of $ {\widetilde \Phi}_{\mathcal D} $.
(The pre-factor $ i $ is removed
for $ {\widetilde \Phi}_{\mathcal D}^{(3)} $ with $ I_{\mathcal D} $.)
In accordance with the up sector,
the hierarchical quark masses and mixings may be reproduced
in terms of $ S^3_{q{\rm L}} $ and $ S^2_{q{\rm L}} $
in Eq. (\ref{S3qL-S2qL}) as
\begin{eqnarray}
| {\tilde \phi}_{bj} | \sim m_b \gg | {\tilde \phi}_{sj} | \sim m_s
\gg | {\tilde \phi}_{dj} | \sim m_d .
\label{phid-tilde}
\end{eqnarray}
The left-handed mixing $ V^{(0)}_{d{\rm L}} $ in Eq. (\ref{Md-Phid})
is taken as the pre-CKM matrix (Particle Data Group convention)
with the vanishing complex phase $ \delta_{13}^{(0)} \simeq 0 $
and the mixing angles $ \theta_{ij}^{(0)} \sim m_{d_i} / m_{d_j} $
($ i < j $) from Eq. (\ref{phid-tilde}).
Then, by including the $ d $-$ D $ mixing effects
as seen below, the CKM matrix with sufficient $ CP $ violation
can be reproduced with reasonable values of $ \theta_{ij}^{(0)} $,
which are adjustable in terms of the USY phases.
The right-handed mixing $ V^{(0)}_{d{\rm R}} $,
on the other hand, may be absorbed practically into $ d_{j{\rm R}} $'s
without physical effects.

The $ d $-$ D $ mixing terms in Eq. (\ref{MD-total}) are given as
\begin{eqnarray}
{\widetilde \Delta}_{dD}
& \simeq & \kappa \left( \begin{array}{ccc}
{\tilde \phi}_{D_1 d} & {\tilde \phi}_{D_1 s} & {\tilde \phi}_{D_1 b} \\
{\tilde \phi}_{D_2 d} & {\tilde \phi}_{D_2 s} & {\tilde \phi}_{D_2 b}
\end{array} \right) ,
\\
{\widetilde \Delta}_{dD}^\prime & \simeq & \left( \begin{array}{cc}
- i {\rm e}^{-i \theta} {\tilde \phi}_{d D_1} & i {\tilde \phi}_{d D_2} \\
- i {\rm e}^{-i \theta} {\tilde \phi}_{s D_1} & i {\tilde \phi}_{s D_2} \\
- i {\rm e}^{-i \theta} {\tilde \phi}_{b D_1}
     & i {\tilde \phi}_{b D_2} + {\sqrt{15}}
\end{array} \right) ,
\end{eqnarray}
where $ {\tilde \phi}_{D_k d_j}
= ( {\widetilde \Phi}_{\mathcal D} )_{3+k,j} $
and $ {\tilde \phi}_{d_i D_k}
= ( {\widetilde \Phi}_{\mathcal D} )_{i,3+k} $.
These $ d $-$ D $ mixing terms provide certain corrections
to $ {\widetilde M}_d $,
which may be evaluated perturbatively as
\begin{eqnarray}
( \delta {\widetilde M}_d )_{ij}
\simeq - \sum_{D_k} ( {\widetilde \Delta}_{dD}^\prime )_{i4}
( {\widetilde \Delta}_{dD} )_{4j} / m_{D_k} .
\label{dMd}
\end{eqnarray}
Then, mainly through $ D_1 $,
significant imaginary parts are provided to $ V_{ub} $ and $ V_{td} $
for the desired $ CP $ violation as
\begin{eqnarray}
{\rm Im} [ V_{ub} ] & \simeq & {\rm Im} [ V_{td} ] \simeq
{\rm Im} [ ( \delta {\widetilde M}_d )_{13} / ( {\widetilde M}_d )_{33} ]
\nonumber \\
& \simeq & {\sqrt{\frac{5}{2}}}
\frac{{\tilde \phi}_{d D_1} {\tilde \phi}_{D_1 b}}{{\tilde \phi}_{b3}}
\frac{\cos \theta}{| \Delta |} \sim - 0.003 .
\label{Vub-Vtd}
\end{eqnarray}

In total, the left-handed mixing $ V_{d{\rm L}} $
for the ordinary $ d_{i{\rm L}} $'s
is determined as the $ 3 \times 3 $ submatrix of the unitary matrix
to diagonalize the entire $ {\widetilde{\mathcal M}}_{\mathcal D} $
in Eq. (\ref{MD-total})
\cite{quark-singlets,CP-EQ1,CP-EQ2,HY}.
Then, the weak charged current mixing matrix $ V $ (CKM matrix)
for the ordinary quarks is given ($ V_{u{\rm L}} \simeq {\bf 1} $) by
\begin{eqnarray}
V = V_{u{\rm L}}^\dagger V_{d{\rm L}} .
\end{eqnarray}
Here, the case of diagonal $ {\widetilde M}_d $ in Eq. (\ref{Md-Phid})
($ V_{d{\rm L}}^{(0)} = V_{d{\rm R}}^{(0)} = {\bf 1} $)
may be specifically interesting,
where the CKM mixing emerges entirely from the $ d $-$ D $ mixing
in the hierarchical basis.
In this case, $ V_{us} $, in particular, is estimated as
\begin{eqnarray}
| V_{us} | \simeq
\frac{| ( \delta {\widetilde M}_d )_{12} |}
{| m_s^{(0)} + ( \delta {\widetilde M}_d )_{22} |}
\lesssim \frac{| V_{ub} |/| V_{cb} |}{| \cos \theta |} ,
\label{Vus}
\end{eqnarray}
where the relations $ | m_s^{(0)} + ( \delta {\widetilde M}_d )_{22} |
\geq | {\rm Im} [ ( \delta {\widetilde M}_d )_{22} ] | $,
and $ | {\tilde \phi}_{d D_1} |/| {\tilde \phi}_{s D_1} | \simeq
| ( \delta {\widetilde M}_d )_{1j} |/| ( \delta {\widetilde M}_d )_{2j} |
\simeq | V_{ub} |/| V_{cb} | $ are considered.

The $ d $-$ D $ mixing also induces small corrections
to the weak neutral currents,
which are related to the unitarity violation of $ V_{d{\rm L}} $
\cite{quark-singlets,CP-EQ1,CP-EQ2,HY}.
We estimate, in particular,
$ | ( V_{d{\rm L}}^\dagger V_{d{\rm L}} )_{33} - 1 |
\simeq | ( V_{d{\rm L}} V_{d{\rm L}}^\dagger )_{33} - 1 | \simeq
| ( {\widetilde \Delta}_{dD}^\prime )_{35} ( \Delta / 5 ) |^2 / m_{D_1}^2
+ | ( {\widetilde \Delta}_{dD}^\prime )_{35} |^2 / m_{D_2}^2
\simeq 3 / \kappa^2 $,
where the correction to $ ( {\widetilde \Delta}_{dD}^\prime )_{34} $
through the $ D_{1{\rm R}} $-$ D_{2{\rm R}} $ mixing
$ \simeq | \Delta / 5 | $ is included.
Then, in order to suppress the correction to $ R_b $
for $ Z \rightarrow b {\bar b} $ to be less than 0.1 \%,
\begin{eqnarray}
\kappa = v_S / v \gtrsim 50
\label{constraint-Delta-k}
\end{eqnarray}
is required, implying $ m_{D_1} \gtrsim 1 {\rm TeV} $
with $ | \Delta | \gtrsim 0.5 $.
This hierarchy of the VEV's may be realized naturally
in some supersymmetric model with an extra gauge symmetry
($ \subset {\rm E6} $) spontaneously broken by $ \langle S \rangle = v_S $.
The quark singlet with $ m_{D_1} \sim 1 {\rm TeV} $
may provide a sizable contribution to the neutron electric dipole moment,
while the effect on $ \epsilon^\prime / \epsilon $ will be small enough
\cite{CP-EQ1,CP-EQ2}.

\begin{figure}[t]
\begin{center}
\scalebox{0.4}{\includegraphics*[7.5cm,4.0cm][17.5cm,18.0cm]{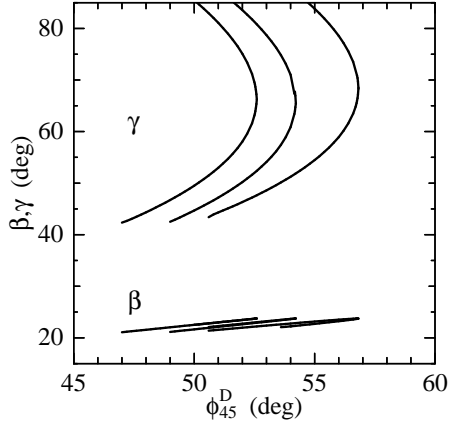}}
\caption{The $ CP $ violation angles $ \beta $ and $ \gamma $
of the CKM matrix versus the USY phase $ \phi^D_{45} $ are shown,
where $  {\tilde \phi}_{d D} $ is taken typically as
$ - 0.003 $ (right), $ - 0.01 $ (left), $ - 0.02 $ (middle).
}
\label{dbg}
\end{center}
\end{figure}

A numerical result is obtained for the CKM matrix
with the $ CP $ violation angles as
\begin{eqnarray}
&& | V | = \left( \begin{array}{ccc}
0.9746 & 0.2240 & 0.0037 \\
0.2239 & 0.9738 & 0.0400 \\
0.0078 & 0.0395 & 0.9986 \end{array} \right) ,
\nonumber \\
&& \alpha = 96.8^\circ , \beta = 23.6^\circ , \gamma = 59.6^\circ ,
\nonumber
\end{eqnarray}
and the rephasing invariant $ CP $ violation measure
$ J = 2.81 \times 10^{-5} $.
The USY phases are taken suitably with $ \kappa = 50 $;
$ \phi^u_{ij}
= 3 [ U_q {\rm diag} ( m_u / m_t , m_c / m_t , 0 ) U_q^\dagger ]_{ij} $
for $ {\widetilde M}_u $ ($ {\widetilde \Phi}_u $) in Eq. (\ref{Mu-Phiu})
with $ V_{u{\rm L}} = V_{u{\rm R}} = {\bf 1} $;
$ \phi^d_{i2} $, $ \phi^d_{i3} $, $ \phi^d_{i4} $ with $ \phi^d_{i1} = 0 $
for $ {\widetilde M}_d $ ($ {\widetilde \Phi}_{\mathcal D}^{(3)} $)
in Eq. (\ref{Md-Phid})
with $ V_{d{\rm L}}^{(0)} = V_{d{\rm R}}^{(0)} = {\bf 1} $
and $ ( m_d^{(0)} / m_d , m_s^{(0)} / m_s , m_b^{(0)} / m_b ) $
= $ ( 1. 026 , 2.067 , 1.008 ) $;
$ \phi^d_{i5} = ( U_q )_{ij} {\tilde \phi}_{d_j D} $
with $ ( {\tilde \phi}_{d D} , {\tilde \phi}_{s D} , {\tilde \phi}_{b D} ) $
= $ ( - 0.01 , -0.107 , 0 ) $ for $ {\widetilde \Delta}_{dD}^\prime $;
$ ( \phi^D_{41} , \phi^D_{42} , \phi^D_{43} , \phi^D_{44} , \phi^D_{45} ) $
= $ ( 0 , 0 , 0.0405 , - 0.0139 , 0.911 ) $
with $ \phi^D_{5J} = 0 $ for $ {\widetilde \Delta}_{dD} $
and $ {\widetilde M}_D $.
The quark masses are obtained as
\begin{eqnarray}
&& m_u = 3 {\rm MeV} , m_c = 1.25 {\rm GeV}, m_t = 177 {\rm GeV} ,
\nonumber \\
&& m_d = 6 {\rm MeV} , m_s = 100 {\rm MeV} , m_b = 4.25 {\rm GeV} ,
\nonumber \\
&& 
m_{D_1} = 1.66 {\rm TeV} , m_{D_2} = 9.18 {\rm TeV}
\nonumber
\end{eqnarray}
with $ m_{D_1} / m_{D_2} \simeq | \Delta | / 5 $ ($ | \Delta | = 0.88 $).
The result of a USY phase space scan is also shown in Fig. \ref{dbg}
for the $ CP $ violation angles $ \beta $ (lower) and $ \gamma $ (upper)
versus the USY phase $ \phi^D_{45} $.
The USY phase values are taken as in the above example.
In particular, $ {\tilde \phi}_{d D} $ is taken typically
as $ - 0.003 $ (right), $ - 0.01 $ (left), $ - 0.02 $ (middle)
with $ {\tilde \phi}_{b D} = 0 $,
by considering $ | {\tilde \phi}_{D_1 b} | \lesssim 0.1 $
in Eq. (\ref{Vub-Vtd})
with $ {\tilde \phi}_{b3} \simeq m_b / ( m_t / 3 ) $,
and $ | {\tilde \phi}_{s D_1} |
\simeq (| V_{cb} |/| V_{ub} |)| {\tilde \phi}_{d D_1} | \lesssim 0.1 $.
Then, $ {\tilde \phi}_{s D} $, $ \phi^D_{43} $, $ \phi^D_{44} $
($ \phi^D_{41} = \phi^D_{42} = 0 $) and $ m_s^{(0)} / m_s $
are adjusted for $ | V_{us} | = 0.224 $,
$ | V_{cb} | = 0.040 $ and $ | V_{ub} | = 0.0037 $.
(The small $ {\tilde \phi}_{b D} $ may be eliminated
by rephasing $ D_{5{\rm R}} $, which is almost compensated
with a slight shift of $ \phi^D_{45} $
by $ {\tilde \phi}_{b D} / {\sqrt 3} \sim {\mbox{some degree}} $.)
We have found suitable USY phase values
to reproduce the quark masses and CKM matrix
with $ \beta \simeq 21^\circ - 24^\circ $
and $ \gamma \simeq 40^\circ - 90^\circ $.
This range of $ \beta $ is really reproduced
by the Particle Data Group convention
with $ \gamma \simeq \delta_{13} $
and $ | V_{us} | $, $ | V_{cb} | $, $ | V_{ub} | $
for $ \theta_{12} $, $ \theta_{23} $, $ \theta_{13} $.
(Solutions are not found
for $ \gamma \lesssim 40^\circ $ or $ \gtrsim 90^\circ $
with our computation algorithm.)
As seen in Fig. \ref{dbg},
$ \phi^D_{45} $ takes the maximal value $ {\bar \phi}^D_{45} $
for given $ {\tilde \phi}_{d D} $,
providing $ \gamma = {\bar \gamma} $.
(We have evaluated $ {\bar \gamma} \simeq 66^\circ - 71^\circ $
and $ {\bar \phi}^D_{45} \simeq 53^\circ - 62^\circ $
for $ - 0.03 \leq {\tilde \phi}_{d D} \leq - 0.002 $.)
This corresponds to the condition
$ m_s^{(0)} + {\rm Re} [( \delta {\widetilde M}_d )_{22} ] = 0 $
in Eq. (\ref{Vus}), specifying
$ {\bar \gamma} \simeq \pi - {\bar \theta}
= ( \pi - {\bar \phi}^D_{45} ) / 2 $,
as verified by calculating $ V_{d_{\rm L}} $ roughly with Eq. (\ref{dMd}).
The mass of the lighter quark singlet is estimated
as $ m_{D_1} \approx 1.6 {\rm TeV} ( \kappa / 50 ) $
for $ \phi^D_{45} \approx 50^\circ $.
These results are valid for the experimentally determined range
of $ | V_{us} | $, $ | V_{cb} | $, $ | V_{ub} | $.

The USY structure may be realized just above the electroweak scale
as in some large extra dimension models
\cite{6D,6D-brane}.
On the other hand, if it is given at a very high unification scale,
the robustness under renormalization group should be considered.
We note that the Yukawa couplings
have the specific structures in the hierarchical basis as
\begin{eqnarray}
{\widetilde \Lambda}_u
= \frac{\lambda}{3} \left( \begin{array}{ccc}
0 & 0 & 0 \\
0 & 0 & 0 \\
0 & 0 & 3  \end{array} \right) ,
{\widetilde \Lambda}_{\mathcal D}
= \frac{\lambda}{3} \left( \begin{array}{ccccc}
0 & 0 & 0 & 0 & 0 \\
0 & 0 & 0 & 0 & 0 \\
0 & 0 & 0 & 0 & {\sqrt{15}} \\
0 & 0 & 0 & * & * \\
0 & 0 & 0 & * & {\sqrt{10}} + * \end{array} \right) .
\nonumber
\end{eqnarray}
Here, ``$ * $" denotes the dominant terms
with the large USY phases for $ m_D $'s and the $ CP $ violation,
while ``0" the perturbation ones with the small USY phases
for the ordinary quark masses, except for $ m_t $, and mixings.
By setting the small USY phases to be zero,
chiral symmetries $ {\rm U(2)}_{q{\rm L}}
\times {\rm U(2)}_{u{\rm R}} \times {\rm U(3)}_{d{\rm R}} $
really appear.
In particular, $ {\rm U(2)}_{q{\rm L}} $ may break as
$ {\rm U(2)}_{q{\rm L}} \rightarrow {\rm U(1)}_{q1{\rm L}}
\rightarrow {\mbox{\rm non}} $
in accordance with  Eq. (\ref{S3qL-S2qL}).
By virtue of these approximate symmetries
the above USY structure is almost maintained
under the renormalization group evolution.
Then, by including the renormalization group corrections,
the suitable USY phase values will be found at the unification scale
in some reasonable range to reproduce the quark masses
and CKM matrix with sufficient $ CP $ violation,
as investigated so far.

In summary, we have investigated the quark masses and mixings
in the USY scheme by including vector-like down-type quark singlets.
In contrast with the standard model with USY,
the sufficient $ CP $ violation is obtained for the CKM matrix
through the mixing between the ordinary down-type quarks and quark singlets.
Two or more quark singlets are needed
to have the relevant large USY phases for the desired $ CP $ violation.
These quark singlets may have masses $ \sim $ TeV,
to be discovered in the future collider experiments
\cite{D-search}.
We have shown that with rather flexible choices of the USY phase values
the actual quark masses and CKM matrix are really reproduced.
Then, it is interesting for further investigations
to invoke some textures and flavor symmetries for the USY phases
so as to derive some predictive relations
among the quark masses and mixings.
The top-bottom hierarchy $ m_t \gg m_b $
also appears naturally in the USY scheme
in the presence of extra down-type quark singlets
but no extra up-type quark singlets.
Furthermore, in the USY scheme (or more generally flavor democracy),
the fermion mass hierarchy may be extended as $ m_t \gg m_b \sim m_\tau $
if vector-like lepton doublets are also present.
In E6-type models, such down-type quark singlets and lepton doublets
are indeed accommodated in the {\bf 27} representation.

\end{document}